\documentclass[12pt]{article}
\usepackage{geometry} 
\usepackage[parfill]{parskip} 
\usepackage{graphicx}

\title{Scattering amplitudes for eight gauge fields}
\author{Andrew Hodges\thanks{ andrew.hodges@wadh.ox.ac.uk, http://www.twistordiagrams.org.uk}\\{\footnotesize  Wadham College, University of Oxford, Oxford OX1 3PN, U.K.}}
\date{March 2006} 

\begin{document}

\maketitle
\begin{abstract}
We study the scattering of eight gauge fields, and give all the tree-level amplitudes  in the helicity-conserved sector. New symmetries are noted, suggesting that significant  further simplification can be achieved.
\end{abstract}
\section{Introduction} 
The calculation of scattering amplitudes for non-Abelian gauge fields has always presented a challenging problem to quantum field theorists, but striking advances have been made in recent years. In an earlier note (Hodges 2005a) it was observed that the powerful on-shell recursion relation found by Britto, Cachazo and Feng (2004b), and proved by Britto, Cachazo, Feng and Witten (2005), could be translated into a simple rule for the composition of twistor diagrams. It was further pointed out (Hodges 2005b) how the twistor diagram formalism could be developed to represent expressions arising from the quadruple-cut method, also  pioneered by Britto, Cachazo and Feng (2004a). An observation of Roiban, Spradlin and Volovich (2004) was then used to define an extension of the earlier twistor diagram composition rule, allowing the amplitudes for all helicities to be treated alike.

It was stated that this approach would effect a saving in computational complexity, by treating all helicity combinations at once, rather than applying the case-by-case analysis pursued hitherto. We now illustrate this statement by using it to give explicit formulas for eight gauge fields in the helicity-conserved sector. The object of this note is limited to giving a complete statement of the amplitudes for every helicity combination, but in tabulating the results it becomes obvious that there is scope for further simplification.  To prepare for these observations we first consider the simpler cases of four and six fields. 

In what follows  we shall consider interactions of $n$ fields with $+$ or $-$ helicity, in a cyclic order given by $(1234\ldots...n)$, and use momentum spinor products indicated by $\langle ij \rangle, [ij]$. Our conventions are such that $ [ij]\langle ij\rangle + [jk]\langle jk\rangle  + [ki]\langle ki \rangle = \frac{1}{2}(i + j + k)^2$ and is abbreviated to $S_{ijk}$  .The rotating, reversing and dualizing operators $g, r$ and $d$ are defined by
\begin{equation}g(1234\ldots n) = (2345\ldots n1),\:\:  r(1234\ldots n)= (n\ldots 4 3 2 1)
\end{equation}
\begin{equation} d(+)=-, d(-)=+,  d\langle\: \rangle= [\:], d[\:]=\langle\: \rangle
\end{equation}
so that
$$d^2=r^2=g^n=1, gr=rg^{-1}, gd=dg, rd=dr \:.$$

\section{Four fields}
The results for four fields are well-known. Their generalization to the `maximum helicity violation' (MHV) sectors of $n$-field scattering has been clear since the work of Parke and Taylor (1986). For four fields, define
\begin{equation}G_{4}(1234)=(\langle12\rangle\langle23\rangle \langle34\rangle\langle41\rangle)^{-1} \delta(\sum_{i=1}^{4}p_i)\:.
\end{equation}
We shall refer in what follows to {\it helicity patterns} such as $(--++)$, and write the amplitudes in the form  $A(1234,H)$, for instance writing $A(1234,--++)$ instead of $A(1^-2^-3^+4^+)$.
The purpose of this notation is to allow the definition of {\it helicity factors}
$$F(--++) = \langle12\rangle, F(-+-+) = \langle13\rangle, F(++--) = \langle34\rangle, \mbox{etc.}$$
and to write the Parke-Taylor amplitude formula in the form:
\begin{equation}A(1234,H) = F(H)^4 G_4(1234)\: .
\end{equation}
The helicity factors are related through the symmetries of $G_4$, which means that there is no need to specify them independently.
The action of $d$ and $r$ on a helicity pattern is given by the obvious definition; for the rotation of a pattern it is convenient to define $g$ so that if $H=(h_1 h_2 \ldots h_n)$ then $gH = (h_n h_1 h_2 \ldots h_{n-1})$. 
With this definition,  $$A(1234,H) =  gA((1234),g^{-1}H) =  rA((1234),rH)\,.$$
Now, because $G_4$ is invariant under the action of $g$ and $r$, we have
$$F(gH) = gF(H), F(rH) = rF(H)$$
and so all the helicity factors can be generated from $F(--++)$ and \mbox{$F(-+-+)$.} 

The helicity factors are only defined up to sign; this is not a problem because in application they will always be raised to the fourth power. Up to sign, they show an extra symmetry which is {\em not} imposed through the symmetry of $G_4$, namely complete symmetry in $(1234)$. 

They have another, non-linear, property which again does not require overall sign. If $a, b$ and $c$ are complex numbers, we may say they {\em form a triangle} iff \begin{eqnarray}&(a+b+c)(a-b-c)(a+b-c)(a-b+c)=0 \nonumber\\ \Leftrightarrow& a^4 + b^4 + c^4 =2(a^2b^2+b^2c^2+c^2a^2)
\end{eqnarray}
With this definition, we note that
$$F(++--)F(--++),\: \:  F(+--+)F(-++-),\:\: F(+-+-)F(-+-+)$$
form a triangle.

The amplitude has a dual formulation:
\begin{equation}A(1234,H)=d\{(F(dH))^4 G_4(1234)\}
\end{equation}
so that, for instance, we have both
\begin{eqnarray}
A(1234,+-+-) =& {\displaystyle \frac{\langle13\rangle^4}{\langle12\rangle\langle23\rangle \langle34\rangle\langle41\rangle}} & \delta(\sum_{i=1}^{4}p_i)\\
\mbox{and }\: A(1234,+-+-)= &{\displaystyle \frac{[24]^4}{[12][23] [34[41]}} & \delta(\sum_{i=1}^{4}p_i)\:.
\end{eqnarray}
The extension of these four-field definitions to all MHV amplitudes is immediate. Obviously this algebraic machinery is unnecessary for describing the MHV formulas, since they are so extraordinarily simple, but it is useful for the non-MHV six-field amplitudes that we shall consider next, and virtually essential for our main goal, the description of the eight-field amplitudes.
\newpage
\section{Six fields}
The  same general pattern of factorization will be found in non-MHV amplitudes. We shall first consider the helicity-conserving interactions of six fields.
Define the singular function
\begin{equation}G_6(123456)=([12][23] \langle45\rangle  \langle56\rangle [1|2+3|4\rangle [3|4+5|6\rangle S_{123})^{-1} \delta(\sum_{i=1}^{6}p_i)\:.
\end{equation}
$G_6$ is {\em not} invariant under $g$ or $r$, but it is invariant under $g^3d$, $g^3r$ and $rd$.

Then the amplitude for six fields, in the helicity-conserved sector, is given by:
\begin{equation}A(123456,H)= (F(H))^4G_6 + g^2\{(F(g^{-2}H))^{4}G_6\} + g^4\{(F(g^{-4}H))^4G_6\}
\end{equation}
and also by a dual formula, which is equivalent to:
\begin{equation}A(123456,H)= g\{F(g^{-1}H))^4G_6\} + g^3\{(F(g^{3}H))^4G_6\} + g^{-1}\{(F(gH))^4G_6\}
\end{equation}
where the helicity factors $F(H)$ will be defined below. These formulas follow from applying the methods given in (Hodges 2005b), which extend the BCF recursion relation and bring all the helicity cases into a single form. The helicity factors can readily be found by these same methods.

The 20 helicity factors, written out by grouped by cycle type for convenient application in these formulas, appear as follows:
\begin{eqnarray}
F(+-+-+-)&=&[13]\langle46\rangle\\
F(-+-+-+)&=&[2|1+3|5\rangle\\
\vspace{0mm} \nonumber\\
F(--+-++)&=&[3|1+2|4\rangle\\
F(-+-++-)&=&[2|1+3|6\rangle\\
F(+-++--)&=&[13]\langle56\rangle\\
F(-++--+)&=&[23]\langle45\rangle\\
F(++--+-)&=&[12]\langle46\rangle\\
F(+--+-+)&=&[1|2+3|5\rangle\\
\vspace{0mm} \nonumber\\
F(++-+--)&=&[12]\langle56\rangle\\
F(+-+--+)&=&[13]\langle45\rangle\\
F(-+--++)&=&[2|1+3|4\rangle\\
F(+--++-)&=&[1|2+3|6\rangle\\
F(--++-+)&=&[3|1+2|5\rangle\\
F(-++-+-)&=&[23]\langle46\rangle\\
\vspace{0mm} \nonumber\\
F(---+++)&=&S_{123}\\
F(--+++-)&=&[3|1+2|6\rangle\\
F(-+++--)&=& [23]\langle56\rangle\\
F(+++---)&=&0\\
F(++---+)&=&[12]\langle45\rangle\\
F(+---++)&=&[1|2+3|4\rangle
\end{eqnarray}
Of these expressions, that for $F(-+-+-+)$ is of particular interest. It cannot be obtained by the recursion formula proved by Britto, Cachazo, Feng and Witten (2005), because of the specific helicity restriction in that formula. It could, however, be obtained by the `quadruple-cut' method, as formulated by Britto, Cachazo and Feng (2004a). In Example 2 below we shall use it to give a new formula for the amplitude in the alternating-helicity case.

{\bf Example 1:}  $A(123456, ---+++)$ is given by:
$$F(---+++)^4 G_6 + g^2 \{F(g^4(---+++))^4 G_6\} +  g^4 \{F(g^2(---+++))^4 G_6\} $$
$$=F(---+++)^4 G_6 + g^2 \{F(-+++--)^4 G_6\} +  g^4 \{F(++---+)^4 G_6\} $$
$$= (S_{123})^4 G_6 + g^2\{ [23]^4 \langle56\rangle^4  G_6\} + g^4\{[12]^4 \langle45\rangle^4  G_6\}$$
\begin{equation}
=
\left(
\begin{array}{lr}
 &\displaystyle{\frac{(S_{123})^3}{[12][23]\langle45\rangle \langle56\rangle [1|2+3|4\rangle  [3|4+5|6\rangle}}\\
 \vspace{0mm}\\
+&
\displaystyle{\frac{\langle12\rangle^3[45]^3}{\langle 16\rangle [34]  [3|4+5|6\rangle [5|6+1|2\rangle S_{612}}}\\
\vspace{0mm}\\
+&
\displaystyle{\frac{\langle23\rangle^3[56]^3}{\langle 34\rangle [16]  [1|2+3|4\rangle [5|6+1|2\rangle S_{234}}}
\end{array}
\right)\: \delta(\sum_{i=1}^{6}p_i)\,.
\end{equation}

By the dual formula, $A(123456, ---+++)$ is given by:
$$g\{F(g^{-1}(---+++))^4 G_6\} + g^3 \{F(g^3(---+++))^4 G_6\} +  g^5 \{F(g(---+++))^4 G_6\} $$
$$=F(--+++-)^4 G_6 + g^3 \{F(+++---)^4 G_6\} +  g^5 \{F(+---++)^4  G_6\} $$
$$=g \{[3|4+5|6\rangle^4 G_6\} + 0 + g^5\{[1|2+3|4\rangle^4 G_6\}$$
\begin{equation}=
\left( \begin{array}{lr}
&\displaystyle{\frac{[4|5+6|1\rangle^3}{ [34][23]\langle56\rangle \langle61\rangle   [2|3+4|5\rangle S_{234}}}\\
\vspace{0mm}\\
 +&
\displaystyle{ \frac{[6|1+2|3\rangle^3}{  [61][12]\langle34\rangle \langle45 \rangle [2|3+4|5\rangle S_{612}}}
\end{array}
\right)\: \delta(\sum_{i=1}^{6}p_i)\,.
\end{equation}

{\bf Example 2:}  $A(123456, +-+-+-)$ is given by:
$$F(+-+-+-)^4 G_6 + g^2\{F(+-+-+-)^4 G_6 \} + g^4\{F(+-+-+-)^4 G_6 \}$$
\begin{equation}
= (1 + g^2 + g^4)\: \: \frac{[13]^4\langle46\rangle^4}{[12][23] \langle45\rangle \langle56\rangle [1|2+3|4\rangle [3|4+5|6\rangle S_{123} }\delta(\sum_{i=1}^{6}p_i)\,.
\end{equation}

By the dual formula,  $A(123456, +-+-+-)$ is also:
$$g\{F(-+-+-+)^4 G_6\} + g^3\{F(-+-+-+)^4 G_6\} + g^5 \{F(-+-+-+)^4 G_6\}$$
\begin{equation}= (1 + g^2 + g^4)\: \:  \frac{[3|2+4|6\rangle^4}{\langle56\rangle\langle61\rangle[23][34] [2|3+4|5\rangle  [4|5+6|1\rangle  S_{234}}\delta(\sum_{i=1}^{6}p_i)\,.
\end{equation}

The equivalence of these formulas for six fields, in the various different helicity cases, is not at all obvious. In the approach we have followed, they can all be regarded as components of a single helicity-independent  `hexagon identity' in the twistor diagram formalism.

It is clear that the 20 helicity factors need not be specified independently. Firstly, the symmetries of $G_6$ mean that it is sufficient to specify just nine of them, the remaining eleven being deducible by the rules
\begin{equation}F(rgH)=rgF(H), F(g^3rH) = g^3rF(H), F(rdH)=rdF(H)
\end{equation}
But there are further symmetries. It is sufficient to specify just the four terms:
\begin{eqnarray}
F(---+++)&=&S_{123}\\
F(+---++)&=&[1|1+2+3|4\rangle\\
F(++---+)&=&[12]\langle45\rangle\\
F(+++---)&=&0
\end{eqnarray}
with the remainder to be obtained by applying symmetry in $(123), (456)$.
Moreover even these four are constrained by thirty triangle-forming identities. Fix any one of the six as a $+$, another as a $-$, then $\pm$ variation over the remaining four yields a non-linear identity analogous to that noted in the four-field case. For example, by fixing $1^+4^-$ we find that
\begin{eqnarray*}&F(+++---)F(+---++), \\
&F(++---+)F(+-+-+-),\\
&F(+-+--+)F(++--+-),
 \end{eqnarray*}
form a triangle. Given this property, it is only necessary to state $F(+---++)$ to derive them all.
It therefore seems very probable that the helicity factors can all be specified by a more economical unifying formula.

\section{Eight fields}
In the helicity-conserving sector for eight fields, we find that three different singular functions are required. Using the same letters as were used by Britto, Cachazo and Feng (2004b) who discovered them, we shall refer to these as $T_8, U_8$ and $V_8$. Indeed the following results simply follow their analysis, but extend it to all helicity patterns rather than restricting attention to $(+-+-+-+-)$. 
Define 
\begin{eqnarray}
T_8(12345678) =&
\left(
\begin{array}{c}
[12][23][78][81]\langle45\rangle\langle56\rangle\\
\mbox{[}1|2+3|4\rangle [1|7+8|6\rangle\\
\mbox{[}1|(2+3)(4+5+6)|7]\\
\mbox{[}3|(4+5+6)(7+8)|1]
  \end{array} 
  \right)^{-1} & \delta(\sum_{i=1}^{8}p_i)\\
V_8(12345678) =&
\left(
\begin{array}{c}
\langle23\rangle \langle34\rangle [56][67]\\
\mbox{[}5|6+7|8\rangle \mbox{[}1|2+3|4\rangle\\
\langle8|(5+6+7)(3+4)|2\rangle\\ \mbox{[}1|(2+3+4)(5+6)|7]\\ S_{234}S_{567}
  \end{array} 
  \right)^{-1}& \delta(\sum_{i=1}^{8}p_i)\\
U_8(12345678)=&
\left(
\begin{array}{c}
\langle23\rangle \langle34\rangle \langle67\rangle \langle78\rangle \\
\mbox{[}5|3+4|2\rangle [1|2+3|4\rangle\\
\mbox{[}5|6+7|8\rangle  \mbox{[}1|7+8|6\rangle\\
 S_{234}S_{567}
  \end{array} 
  \right)^{-1} & \delta(\sum_{i=1}^{8}p_i)\:.
\end{eqnarray}

$T_8$ is invariant under $gr$, $V_8$ is invariant under $rd$, $U_{8}$ is invariant under $gr$ and $g^4$. These symmetries are made manifest in the twistor diagram representations of (Hodges 2005b).
\newpage
We can now give the general 20-term formula for the amplitude $A(12345678,H)$ explicitly in terms of the momentum spinors because the helicity factors have all been tabulated and will be found in  \S6 below. The formula is:
\begin{equation} \sum_{i=0}^7 (gd)^i (F_{T}((gd)^{-i} H))^4 T_8 +  \sum_{i=0}^7 (gd)^i (F_{V}((gd)^{-i} H))^4 V_8 +  \sum_{i=0}^3 (gd)^i (F_{U}((gd)^{-i} H))^4 U_8
\end{equation}
and the dual formula is:
\begin{equation} d\left\{ \sum_{i=0}^7 (gd)^i (F_{T}((gd)^{-i} dH))^4 T_8 +  \sum_{i=0}^7 (gd)^i (F_{V}((gd)^{-i} dH))^4 V_8 +  \sum_{i=0}^3 (gd)^i (F_{U}((gd)^{-i} dH))^4 U_8 \right\}
\end{equation}
{\bf Example:} For $A(12345678, +-+-+-+-)$ the first formula yields the expression found by Britto, Cachazo and Feng (2004b), viz.:
\begin{equation}  \sum_{i=0}^7 (gd)^i [13]^4 [17]^4 \langle46\rangle^4 T_8
+  \sum_{i=0}^7 (gd)^i \langle24\rangle^4 [57]^4 [1|2+3+4|8\rangle^4 V_8
+  \sum_{i=0}^3 (gd)^i [15]^4 \langle24\rangle^4 \langle68\rangle^4 U_8
\end{equation}
The dual formula, however, gives a new expression:
\begin{eqnarray}
 \sum_{i=0}^7 (gd)^i (\langle28\rangle [5|2+3|1\rangle + \langle21\rangle [5|4+6|8\rangle)^4 dT_8 \nonumber\\
 +  \sum_{i=0}^7 (gd)^i [3|(2+4)(1+2+3+4)(5+7)|6\rangle^4 dV_8 \nonumber \\
 +  \sum_{i=0}^3 (gd)^i [3|(2+4)(6+8)|7]^4 dU_8 
\end{eqnarray}

\section{Remarks}
It must be emphasised that the validity of the amplitude formulas thus found depends entirely on the powerful field-theoretic discoveries due to Britto, Cachazo, Feng and others. However, the twistor diagram formalism has indeed simplified the systematic extension of their results to all helicity patterns. 

The factorization of the amplitudes into  a singular function and a numerator helicity factor reflects the way that the helicity-independent twistor diagram approach separates the two different sources of complexity. The singular function corresponds to the geometric structure defined by the vertices and edges of the  twistor diagram (the `quilt').  This has highly non-trivial  features,  giving rise to multiple representations of the amplitude, but the formalism simplifies these by making them independent of helicity pattern. 

The helicity factor then corresponds to the homogeneities of the external fields in their twistor representation. The list  in $\S6$ gives the factors explicitly.
The list may appear long, but most of the helicity factors can be written down immediately from the diagram formulation, and the remainder need little more than spinor algebra and the binomial theorem --- all pencil-and-paper work. So the formalism we have developed has, as intended, reduced the computational complexity. 

But it will be apparent that the list as given below is redundant; as in the six-field case there are further symmetries, which are obscured when the factors are listed by cycle type. There is  every reason to suppose that a more economical method can be given to specify all helicity factors. In particular, it appears very likely that  supersymmetric algebra will yield a concise formula and show how the symmetries arise and why the Parke-Taylor factors generalise as they do. Our object would be to reduce this helicity factor aspect to near-triviality, and then focus on the much deeper question of the representation of amplitudes (now both colour-stripped and helicity-stripped) by the helicity-independent twistor diagrams of (Hodges 2005b).

The restriction to the `helicity-conserved sector' for eight fields has been stated throughout, but this is not a significant restriction: a similar analysis can be applied to the remaining sectors. Further details of the twistor-diagram methods employed will be given  in later papers. 

\section{Eight-field helicity factors}
There are 70 helicity patterns in the helicity-conserved sector. The corresponding factors are listed below, grouped by cycle type, as is convenient for applying in the amplitude formula, but rather than listing all 70 we have tabulated only a smaller number, from which the remainder can readily be derived by exploiting the symmetries of the relevant singular function. Each is only defined up to sign.

In each case it will readily be observed that as in the six-field case there are additional symmetries. Less obvious is that  for each of the $T_8, V_8, U_8,$ there are 420 triangle-forming identities satisfied by the factors. Each factor participates in 36 of these non-linear identities, creating a super-sudoku of consistency. 
\newpage
{\bf $T_8$ function:}
Only 38 factors are given as the remaining 32 can be deduced from $$F_T(grH)=grF_T(H)\:.$$
{\em Type 1: special symmetry}
\begin{eqnarray}
F_{T}(+-+-+-+-)&=&[13][17]\langle46\rangle \\
F_{T}(-+-+-+-+)&=&[28]  [1|2+3|5\rangle +[21]  [8|4+6|5\rangle\\
F_{T}(++--++--)&=&[12] [1|7+8|4\rangle\\
F_{T}(-++--++-)&=&[23][17]\langle45\rangle\\
F_{T}(++++----)&=&0\\
F_{T}(+++----+)&=&0\\
F_{T}(----++++)&=&[78] [1|2+3|4\rangle\\
F_{T}(---++++-)&=&[1|(2+3)(4+5+6)|7]
\end{eqnarray}
{\em Type 2: $(+++--+--)$}
\begin{eqnarray}
F_{T}(+++--+--)&=&0\\
F_{T}(++--+--+)&=&[12][18]\langle46\rangle\\
F_{T}(--+--+++)&=&[13][78]\langle45\rangle\\
F_{T}(-+--+++-)&=&[12] [7|5+6|4\rangle +[72] [1|2+3|4\rangle\\
F_{T}(+--+++--)&=&[1|(2+3)(7+8)|1]
\end{eqnarray}
{\em Dual of type 2:}
\begin{eqnarray}
F_{T}(++-++---)&=&[12] [1|7+8|6\rangle \\
F_{T}(+-++---+)&=&[13][18]\langle56\rangle\\
F_{T}(-++---++)&=&0\\
F_{T}(---++-++)&=&[78] [1|2+3|6\rangle\\
F_{T}(--++-++-)&=&[17][3|4+6|5\rangle +[37] [1|7+8|5\rangle
\end{eqnarray}
{\em Type 3: $(++-+--+-)$}
\begin{eqnarray}
F_{T}(++-+--+-)&=&[12][17]\langle56\rangle\\
F_{T}(-+--+-++)&=&[12][78]\langle46\rangle\\
F_{T}(+--+-++-)&=&[17] [1|2+3|5\rangle \\
F_{T}(--+-++-+)&=&[38][1|7+8|4\rangle +[18] [3|5+6|4\rangle
\end{eqnarray}
\newpage
{\em Type 4: $(+++-+---)$}
\begin{eqnarray}
F_{T}(+++-+---)&=&0\\
F_{T}(++-+---+)&=&[12][18]\langle56\rangle\\
F_{T}(+-+---++)&=&0\\
F_{T}(-+---+++)&=&[12][78]\langle45\rangle\\
F_{T}(+---+++-)&=&[17][1|2+3|4\rangle\\
F_{T}(---+++-+)&=&[1|(2+3)(4+5+6)|8] \\
F_{T}(--+++-+-)&=&[17] [3|4+5|6\rangle +[37] [1|7+8|6\rangle \\
F_{T}(-+++-+--)&=&[23][1|7+8|5\rangle
\end{eqnarray}
{\em Type 5: $(++--+-+-)$}
\begin{eqnarray}
F_{T}(++--+-+-)&=&[12][17]\langle46\rangle\\
F_{T}(+--+-+-+)&=&[18][1|2+3|5\rangle\\
F_{T}(--+-+-++)&=&[13][78]\langle46\rangle\\
F_{T}(-+-+-++-)&=&[27] [1|2+3|5\rangle +[21] [7|4+6|5\rangle\\
F_{T}(+-+-++--)&=&[13][1|7+8|4\rangle\\
F_{T}(-+-++--+)&=&[12] [8|4+5|6\rangle +[82][1|2+3|6\rangle \\
F_{T}(+-++--+-)&=&[13][17]\langle56\rangle\\
F_{T}(-++--+-+)&=&[18][23]\langle45\rangle
\end{eqnarray}

{\bf $V_8$ function:}
Only 43 factors are given, as the remaining 27 can be derived from them using 
$$F_V(drH)=drF_V(H)\: .$$
{\em Type 1: special symmetry}
\begin{eqnarray}
F_{V}(+-+-+-+-)&=&\langle24\rangle[57][1|2+3+4|8\rangle\\
F_{V}(-+-+-+-+)&=&\langle3|(2+4)(1+2+3+4)(5+7)|6]\\
F_{V}(++--++--)&=&\langle34\rangle[56][1|2+3+4|8\rangle\\
F_{V}(+--++--+)&=&\langle23\rangle[1|(2+3+4)(6+7)|5]\\
F_{V}(--++--++)&=&\langle23\rangle[67]S_{1234}\\
F_{V}(++++----)&=&0\\
F_{V}(+++----+)&=&[1|2+3|4\rangle S_{567}\\
F_{V}(++----++)&=&\langle34\rangle [1|(2+3+4)(5+6)|7]\\
F_{V}(+----+++)&=&0\\
F_{V}(----++++)&=&0
\end{eqnarray}
{\em Type 2: $(+++--+--)$}
\begin{eqnarray}
F_{V}(+++--+--)&=&[1|2+3|4\rangle [6|5+7|8\rangle\\
F_{V}(++--+--+)&=&\langle34\rangle [1|(2+3+4)(6+7)|5]\\
F_{V}(+--+--++)&=&\langle23\rangle [1|(2+3+4)(5+6)|7]\\
F_{V}(--+--+++)&=&\langle24\rangle[67]S_{1234}\\
F_{V}(-+--+++-)&=&0\\
F_{V}(+--+++--)&=&\langle23\rangle[56][1|2+3+4|8\rangle\\
F_{V}(--+++--+)&=& [5|(6+7)(1+2+3+4)(3+4)|2\rangle\\
F_{V}(-+++--+-)&=&  [7|5+6|8\rangle S_{234}
\end{eqnarray}
{\em Type 3: $(+--+--+-)$}
\begin{eqnarray}
F_{V}(++-+--+-)&=&[1|2+4|3\rangle[7|5+6|8\rangle\\
F_{V}(+-+--+-+)&=&\langle24\rangle[1|(2+3+4)(5+7)|6]\\
F_{V}(-+--+-++)&=&\langle34\rangle[57]S_{1234}\\
F_{V}(+--+-++-)&=&\langle23\rangle[67][1|2+3+4|8\rangle \\
F_{V}(-++-+--+)&=&\makebox[2.5 in][l]{$[5|(6+7)(1+2+3+4)(2+3)|4\rangle$}
\end{eqnarray}
{\em Type 4: $(+++-+---)$}
\begin{eqnarray}
F_{V}(+++-+---)&=&[1|2+3|4\rangle[5|6+7|8\rangle\\
F_{V}(++-+---+)&=& [1|2+4|3\rangle S_{567}\\
F_{V}(+-+---++)&=&\langle24\rangle[1|(2+3+4)(5+6)|7]\\
F_{V}(-+---+++)&=&\langle34\rangle[67]S_{1234}\\
F_{V}(+---+++-)&=&\makebox[2.5 in][l]{$0$}
\end{eqnarray}
{\em Type 4, reversed:}
\begin{eqnarray}
F_{V}(---+-+++)&=&\langle23\rangle[67]S_{1234}\\
F_{V}(--+-+++-)&=&\makebox[2.5 in][l]{$0$}\\
F_{V}(-+-+++--)&=&[56]\langle3|(2+4)(5+6+7)|8\rangle\\
F_{V}(+-+++---)&=&[1|3+4|2\rangle[5|6+7|8\rangle\\
F_{V}(-+++---+)&=&S_{234}S_{567}
\end{eqnarray}
{\em Type 5: $(++--+-+-)$}
\begin{eqnarray}
F_{V}(++--+-+-)&=&\langle34\rangle[57][1|2+3+4|8\rangle\\
F_{V}(+--+-+-+)&=&\langle23\rangle [1|(2+3+4)(5+7)|6]\\
F_{V}(--+-+-++)&=&\langle24\rangle[57]S_{1234}\\
F_{V}(-+-++--+)&=&\makebox[2.4 in][l]{$[5|(6+7)(1+2+3+4)(2+4)|3\rangle$}\\
F_{V}(+-++--+-)&=&[1|3+4|2\rangle [7|5+6|8\rangle
\end{eqnarray}
{\em Type 5, reversed:}
\begin{eqnarray}
F_{V}(-+-+--++)&=&[7|(5+6)(1+2+3+4)(2+4)|3\rangle\\
F_{V}(+-+--++-)&=&\langle24\rangle [67] [1|2+3+4|8\rangle\\
F_{V}(-+--++-+)&=&\langle34\rangle[56]S_{1234}\\
F_{V}(-++-+-+-)&=&[57]\langle4|(2+3)(5+6+7)|8\rangle\\
F_{V}(++-+-+--)&=&[1|2+4|3\rangle[6|5+7|8\rangle
\end{eqnarray}
{\bf $U_8$ function:}
Only 23 factors are given, as the remaining 47 can be derived from them using 
$$F_U(grH)=grF_U(H),\: F_U(g^4H)=g^4F_U(H)\: .$$
{\em Type 1: special symmetry}
\begin{eqnarray}
F_{U}(+-+-+-+-)&=& [15]\langle24\rangle\langle68\rangle\\
F_{U}(-+-+-+-+)&=&\langle3|(2+4)(6+8)|7\rangle\\
F_{U}(++--++--)&=&[15]\langle34\rangle\langle78\rangle\\
F_{U}(-++--++-)&=&\langle4|(2+3)(6+7)|8\rangle\\
F_{U}(++++----)&=&\makebox[1.5 in][l]{$0$}\\
F_{U}(+++----+)&=&\langle67\rangle[1|2+3|4\rangle
\end{eqnarray}
{\em Type 2: $(+++--+--)$}
\begin{eqnarray}
F_{U}(+++--+--)&=&\langle78\rangle[1|2+3|4\rangle\\
F_{U}(++--+--+)&=&[15]\langle34\rangle\langle67\rangle\\
F_{U}(--+--+++)&=&\makebox[1.5 in][l]{$\langle24\rangle S_{567}$}
\end{eqnarray}
{\em Dual of type 2:}
\begin{eqnarray}
F_{U}(++-++---)&=&\makebox[1.5 in][l]{$0$}\\
F_{U}(+-++---+)&=&\langle67\rangle[1|3+4|2\rangle\\
F_{U}(-++---++)&=&\langle6|(7+8)(2+3)|4\rangle
\end{eqnarray}
{\em Type 3: $(++-+--+-)$}
\begin{eqnarray}
F_{U}(++-+--+-)&=&\makebox[1.5 in][l]{$\langle68\rangle[1|2+4|3\rangle$}\\
F_{U}(-+--+-++)&=&\langle34\rangle[5|7+8|6\rangle\\
F_{U}(+--+-++-)&=&\langle23\rangle[1|6+7|8\rangle
\end{eqnarray}
{\em Type 4: $(+++-+---)$}
\begin{eqnarray}
F_{U}(+++-+---)&=&\makebox[1.5 in][l]{$0$}\\
F_{U}(++-+---+)&=&\langle67\rangle[1|2+4|3\rangle\\
F_{U}(+-+---++)&=&\langle24\rangle[1|7+8|6\rangle\\
F_{U}(-+---+++)&=&\langle34\rangle S_{567}
\end{eqnarray}
{\em Type 5: $(++--+-+-)$}
\begin{eqnarray}
F_{U}(++--+-+-)&=&[15]\langle34\rangle\langle68\rangle\\
F_{U}(+--+-+-+)&=&\langle23\rangle[1|6+8|7\rangle\\
F_{U}(--+-+-++)&=&\langle24\rangle[5|7+8|6\rangle\\
F_{U}(-+-+-++-)&=&\makebox[1.5 in][l]{$\langle3|(2+4)(6+7)|8\rangle$}
\end{eqnarray}

\section{References}
R. Britto, F. Cachazo and B. Feng, Generalized unitarity and one-loop amplitudes in N=4 superYang-Mills,  hep-th/0412103 (2004a)

R. Britto, F. Cachazo and B. Feng, Recursion relations for tree amplitudes of gluons,  hep-th/0412308 (2004b)

R. Britto, F. Cachazo, B. Feng and E. Witten, Direct proof of tree-level recursion relation in YangMills theory,  hep-th/0501052  (2005) 

A. Hodges, Twistor diagram recursion for all gauge-theoretic tree amplitudes,  hep-th/0503060  (2005a)

A. Hodges, Twistor diagrams for all tree amplitudes in gauge theory: a helicity-independent formalism, hep-th/0512336 (2005b)

S. Parke and T. Taylor, An amplitude for N gluon scattering, Phys. Rev. Lett. {\bf 56,} 2459 (1986)

R. Roiban, M. Spradlin and A. Volovich, Dissolving N=4 amplitudes into QCD tree amplitudes,   hep-th/0412265 (2004)

\end{document}